\documentclass[11pt,a4paper]{emulateapj} 

\usepackage{epsfig}
\usepackage{natbib}

\usepackage{ifpdf} \usepackage[bookmarks=false]{hyperref} \pdfoutput=1
\usepackage[usenames,dvipsnames]{color} \usepackage{graphicx}
\usepackage{graphics} \usepackage{float}
\usepackage{amsmath} \usepackage{amssymb} \usepackage{acronym}
\usepackage{comment} \usepackage{multirow} \usepackage{xspace}
\usepackage{units} \usepackage{dcolumn} \usepackage{ulem}

\newcommand{\thpara}{\boldsymbol{\theta}}
\newcommand{\chmass}{\mathcal{M}_c}

\begin{document}
\title{Basic Parameter estimation of Binary Neutron Star Systems by
  the Advanced LIGO/Virgo Network} 
  \author{Carl L. Rodriguez\altaffilmark{1},
Benjamin Farr\altaffilmark{1},
Vivien Raymond\altaffilmark{1,2}, Will M. Farr\altaffilmark{1},
Tyson B. Littenberg\altaffilmark{1},
Diego Fazi\altaffilmark{1}, \&
Vicky Kalogera\altaffilmark{1}}

\altaffiltext{1}{Center for Interdisciplinary Exploration and Research
  in Astrophysics (CIERA) \& Dept.~of Physics and Astronomy,
  Northwestern University, 2145 Sheridan Rd, Evanston, IL 60208, USA;
  [e-mail: {\tt cr@u.northwestern.edu}]} \altaffiltext{2}{California
  Institute of Technology, Pasadena, CA 91125, USA}

\begin{abstract}

Within the next five years, it is expected that the Advanced LIGO/Virgo network will have
reached a sensitivity sufficient to enable the routine detection of
gravitational waves.  Beyond the initial detection, the scientific
promise of these instruments relies on the effectiveness of our physical
parameter estimation capabilities. A major part of this effort has been towards the
detection and characterization of gravitational waves from compact
binary coalescence, e.g. the coalescence of binary neutron stars.
While several previous studies have investigated the accuracy of parameter
estimation with advanced detectors, the majority have relied
on approximation techniques such as the Fisher Matrix which are insensitive to the non-Gaussian
nature of the gravitational-wave posterior distribution function.  Here we report average
statistical uncertainties that will be achievable for strong detection candidates 
 ($\text{SNR}=20$) over a comprehensive sample of source parameters.  We use
the Markov-Chain Monte Carlo based parameter estimation software developed by the LIGO/Virgo
Collaboration with the goal of updating the previously quoted Fisher Matrix bounds. 
 We find the recovery of the individual masses to be
fractionally within 9\% (15\%) at the 68\% (95\%) credible intervals for equal-mass
systems, and within 1.9\% (3.7\%) for unequal-mass systems.  We also find that the Advanced
LIGO/Virgo network will constrain the locations of binary neutron star mergers to a median
 uncertainty of $5.1$ $\mathrm{deg}^2$ (13.5 $\mathrm{deg}^2$) on the sky.  This region is improved to  2.3 $\mathrm{deg}^2$ $(6$ $\mathrm{deg}^2)$ with the addition of the proposed LIGO India detector to the network.  We also report the
average uncertainties on the luminosity distances and
orbital inclinations of strong detections that can be
achieved by different network configurations.
\end{abstract}

\maketitle
\section{Introduction}

Within the next few years, the first generation of gravitational-wave (GW) interferometers capable of regularly detecting astrophysical sources will
come online \citep{AdvLIGO,AdvVirgo}.  The Advanced LIGO and Advanced Virgo
detectors (and the anticipated LIGO India detector)
 will provide the first insights into the final moments of
 compact object mergers, including the mergers of binary neutron
star systems.  Intense preparations are underway to characterize and
extract as much physical information as possible from these signals.

The mergers of binary neutron stars (BNS) are expected to be one of the most
common compact binary sources in the advanced detector era.  
  Models from stellar
evolution and observations of binary pulsars suggest that the number
of BNS mergers within Advanced LIGO/Virgo's detection
horizon could reach several tens to hundreds each year
\citep{RatesPaper}.  Although the peak sensitivity of ground-based
detectors is not focused on the frequency at which BNS systems merge,
it could still be possible to extract information about both strong
field gravitational physics \citep{Li2012} and the the equation of state of
dense nuclear matter
\citep{HindererBNS2010}.  Furthermore, the observations of multiple BNS systems
will provide key insights into the evolution of binary systems in the
field \citep{VickyRates,KimRates,OsowskiRates2011,RichardRates2010} . As such, BNS systems will likely
form the ``bread and butter'' of the compact binary coalescence
detection effort in the coming years.

Of course one must distinguish between the detection of such events
and the precision measurement of their relevant physical parameters.
The detection of BNS systems will be performed with a grid-based matched
filtering approach.  By comparing the data stream with a bank of
theoretical templates, the time-series data can be searched for
candidate signals \citep{S6search}.
The parameter space
of these signals exhibits degeneracies and strong correlations.  In order
to completely realize the science potential of LIGO and Virgo observations we must
perform a full exploration of the parameter space for each detection.   
 To do so, we characterize the posterior distribution function using Bayesian inference 
to make informative, scientifically meaningful statements about the physics of 
BNS systems.

In this paper we will study large signal to noise ratio BNS coalescence events and report on the measurement capabilities of the LIGO/Virgo advanced detector network.  The particular inference method used in our work is a
Markov chain Monte Carlo (MCMC) sampling code,
\texttt{lalinference\_mcmc}, included in the LIGO Algorithm Library
parameter estimation library LALInference.

There is a long history of research which has sought to provide insight into what will be learned by gravitational wave observations.
The majority of studies have employed the Fisher matrix
formalism which was first adapted for gravitational-wave parameter
estimation by \citet{FinnDetection}.  While each of these studies
\citep{PoissonWill,CutlerFlanagan,ArunPE} have pointed out the
limitations of the Fisher Information Matrix estimates, there have been
relatively few studies which investigate the BNS parameter estimation capabilities
of Advanced LIGO/Virgo using the full infrastructure of Bayesian inference which must be
employed to draw robust conclusions about the GW source and its properties \citep{Vallisneri, Inadequacies}
.

Fisher matrix estimates, which our study seeks to improve upon, approximate the likelihood distribution as a multivariate Gaussian and are thus insensitive to more complicated structures such as multiple peaks, non-trivial correlations, skew, etc.  
Conversely, studies in gravitational wave parameter estimation have repeatedly shown that the recovered posterior distribution functions exhibit the very details
which violate the explicit assumptions of the Fisher matrix formalism. Furthermore, measured quantities which are not tightly constrained, particularly distance and mass ratio, will be influenced by our choice of prior -- a consideration which is typically left out of Fisher-based studies.

By relaxing assumptions about the shape of the likelihood distribution, and therefore the posterior, we provide more detailed insight into the average measurement accuracy for LIGO/Virgo observations of ``loud'' BNS mergers.  These findings will help guide the community as it prepares to employ GW observations as a new tool to study astrophysics, relativity, and cosmology.

he goal of our study is to provide touchstone estimates of parameter estimation uncertainties 
for Advanced LIGO/Virgo observations of strong BNS signals, accounting for the complexity of the likelihood distribution. For ease of accessibility, the results are collected and summarized in 
 Tables \ref{ciTableIntrinsic} and \ref{ciTableExtrinsic}.

 Our study provides complementary results to previous investigations which used similar methods but pursued different questions.
 While our work reports on the recovery of all system parameters for a range of plausible BNS mass and mass-ratios at a fixed (and comparatively large) signal to noise ratio, 
 recent papers 
\citep{Nissanke2011,Nissanke2013}
have prioritized multi messenger astronomy by simulating a LIGO/Virgo detection catalog and strictly reporting on how well  BNS mergers can be localized on the sky and/or in volume to facilitate electromagnetic observations.  We are in agreement with the published work where our findings overlap, and proceed to also examine how well intrinsic quantities, in particular the mass parameters, are measured for strong detections.  Our work also distinguishes itself both from 
\cite{Nissanke2011,Nissanke2013}
and
\cite{Veitch2012}
by quantifying the effects of mass ratio on parameter recovery, whereas the previous studies have restricted their results for BNS systems to the equal mass case.

In Section \ref{PEsection}, we describe the infrastructure of the parameter estimation
code, \texttt{LALInference}, and its associated MCMC sampler,
\texttt{lalinference\_mcmc}, as well as the frequency-domain
gravitational-wave template we employ.  In Section
\ref{resultsSection}, we qualitatively analyze the posterior
probability density functions for BNS systems with different masses
and extrinsic parameters.  We select three equal-mass and one unequal-mass
 binary systems as prototypical examples of BNS systems.  Each system is analyzed 40 
 times with isotropically selected sky locations and orbital orientations, and with a distance
 such that each signal was injected with a network signal-to-noise ratio of 20.  
The results are divided into three subsections of interest: the recovery of the mass
parameters (Section \ref{massSection}), the recovery of the orbital
inclination and luminosity distance (Section \ref{idSection}), and the
localization of sources on the sky (Section \ref{skySection}).
Finally we provide quantitative 1-dimensional credible intervals on
each parameter in Section \ref{ciSection}.  
Throughout this paper we adopt geometrized units with $G=c=1$.

\section{Parameter Estimation}
\label{PEsection}

The parameter estimation methodology used here -- namely matched filtering with post Newtonian waveforms and using MCMC to sample the posterior -- have become sufficiently ubiquitous in the GW literature that we will give only a very cursory treatment of both mainly for the sake of introducing notation and terminology.  We refer readers seeking more detail to Appendix \ref{appendixMCMC} and references therein.

We begin by introducing the matched filtering formalism for parameter estimation.
We assume that the time-domain signal in a gravitational-wave network
can be written as the sum of a gravitational waveform $h_0$ and
the noise of the detector $n$.  We further assume that this noise is
stationary and Gaussian with zero mean.  The detector output is simply
\begin{equation}
s = n + h_0 .
\label{SignalAddition}
\end{equation}
Since the noise model is Gaussian, we can write the probability of a
specific signal realization $s$ given an input waveform $h(\thpara)$ as
proportional to the probability that the residual is Gaussian
distributed once the waveform has been subtracted
\begin{align}
  p(s | \thpara) &\propto \exp\left[-\frac{1}{2}\left<n|n
    \right>\right] \nonumber \\ &= \exp\left[-\frac{1}{2}\left < s -
    h(\thpara) | s-h(\thpara)\right >\right] ,
  \label{likelihood}
\end{align}
where $\thpara$ is the set of parameters for the template waveforms.
The quantity $p(s | \thpara)$ is the likelihood of the signal $s$
given the parameters $\thpara$.  The inner product, $\left< ~|~
\right> $, is defined using the noise spectrum of the detectors as
\begin{equation}
  \left<a|b\right> \equiv 4 \Re \int
  \frac{\tilde{a}(f)\tilde{b}^*(f)}{S_n(f)} df ,
  \label{innerProduct}
\end{equation}
where $S_n(f)$ is the one-sided power-spectral density (PSD) as a
function of frequency, and $\tilde{a}(f)$ and $\tilde{b}(f)$ are the
Fourier transforms of the time-domain data $a(t)$ and $b(t)$.  If we
pick a set of parameters $\thpara$ such that $h(\thpara) = h_0$, then
the likelihood \eqref{likelihood} will be near a global maximum;
however, the presence of noise will in general deflect the maximum of
our likeliood away from the value at $h(\thpara) = h_0$.  Therefore, the
maximum likelihood parameters do not necessarily correspond to the true
parameters of the source.

Once we have the likelihood of the signal \eqref{likelihood}, we
employ Bayes' Theorem to obtain the posterior probability of the system
parameters $\thpara$ given the signal $s$ as
\begin{align}
  p(\thpara | s) &= \frac{p(\thpara)p(s | \thpara)}{p(s)}
  \nonumber\\ & \propto p(\thpara) \exp\left[-\frac{1}{2}\big < s -
    h(\thpara) | s-h(\thpara) \big > \right] ,
  \label{posterior}
\end{align}
where $p(\thpara)$ are the prior probabilities on our source
parameters and $p(s)$ is a normalization constant.

\subsection{Markov-Chain Monte Carlo}
\label{MCMCSection}

The LIGO Algorithm Library Bayesian inference code,
\texttt{LALInference}, is designed as a unified framework for
gravitational-wave parameter estimation.  By using a common setup for
waveform generation, PSD estimation, data handling, and other
associated techniques from gravitational-wave parameter estimation,
\texttt{LALInference} allows the implementation of multiple samplers
of the parameter space, including Nested Sampling
(\texttt{lalinference\_nest}, described in \cite{nestedsampling2010})
and Markov-Chain Monte Carlo (\texttt{lalinference\_mcmc}).  We elect
to use the MCMC sampler for this study.  \texttt{lalinference\_mcmc}
is based upon the previously described code, \texttt{SpinSpiral}
\citep{spinspiral2009, spinspiral2010}.  The MCMC employs a
Metropolis-Hastings sampling algorithm \citep{Gilks99}, which is described in 
Appendix \ref{appendixMCMC}.

\subsection{Parameter Space}
\label{parameterSection}

Our methodology is predicated on the existence of accurate templates.  
We assume no systematic difference between our model waveforms and the true GW signal.
We use a frequency-domain waveform accurate up to 3.5 post-Newtonian (pN) order in phase and 3 pN order in amplitude of the lowest ($l = |m| = 2$) spatial mode. We restrict ourselves to quasi-circular, non-spinning waveforms as a simplifying assumption. The standard form of our waveform model, known as the TaylorF2 approximant, is calculated via the stationary-phase approximation~\mbox{
\citep{BuonannoWaveform}}
.  Future studies will be needed to include the impact of realistic NS spin and/or the bias introduced by using approximate waveforms.

In the absence of any spin, the phase evolution of the gravitational-waveform is determined by four intrinsic parameters -- the two masses ($M_1$ and $M_2$) and two intrinsic constants of integration ($\phi_0$ and $t_c$) .
 In addition, there are 5 extrinsic
parameters which do not influence the inspiral of the binary, but govern
 the amplitude of the signal in each
detector.  Considering these leads to a 9-dimensional parameter space
for non-spinning systems as employed in our MCMC:
\begin{equation}
\thpara = (\chmass, q, \phi_0,t_c,D,\iota,\psi,\alpha,\delta),
\label{parameterspace}
\end{equation}
where
\begin{itemize}
\item $\chmass \equiv (M_1 M_2)^{3/5}  M_{tot}^{-1/5}$ is the chirp mass,
\item $q\equiv M_2 / M_1$ is the mass ratio,
\item $\phi_0$ and $t_c$ are the chirp phase and chirp time, arbitrary
  phasing parameters,
\item $D$ is the luminosity distance to the binary,
\item $\iota$ is the orbital inclination (the angle between the
  orbital angular momentum and the line of sight),
\item $\psi$ is the gravitational-wave polarization, and
\item $\alpha$ and $\delta$ are the right ascension and declination of
  the source on the sky.
\end{itemize}
  Since the wave amplitude depends on the orientation of the binary
  with respect to each detector, most of the information about these
  extrinsic parameters comes from two sources: the time-of-arrival
  triangulation of the signal, and the relative amplitudes
  in each detector in the network.

\subsection{Priors}
We are interested in the posterior $p(\thpara | s)$ as it encodes 
information about both the prior state of knowledge about the problem in
addition to the likelihood of the signal.  We adopt the same conventions for the prior distribution as were used in the follow-up parameter estimation studies of simulated detections during LIGO's sixth, and Virgo's second, science collection periods~\mbox{
\citep{S6PE}}.  They are:
\begin{itemize}
\item uniform in component masses from $0.8M_{\odot} \leq M_{1,2} \leq
  30M_{\odot}$, with a minimum chirp mass 
  of 0.6 $M_{\odot}$ 
\item uniform in volume, which implies a luminosity distance prior of
  $p(D)dD \propto D^2 dD$,
\item uniform in coalescence time over the segment of data being
  analyzed,
\item isotropic in all angles
\end{itemize}
\subsection{Detector Configuration and Noise Models}
  \label{detectorSection}

To perform the integral defined in \eqref{innerProduct}, we used as
our power-spectral density the best estimate for a high-power,
zero-detuning configuration of Advanced LIGO, provided by the LIGO
Scientific Collaboration.  Both the noise curve and technical reports
describing it can be found in \cite{ADVLIGONoise}.  We consider two
configurations of the advanced detector network: a three-detector
configuration consisting of the two LIGO sites (in Hanford, WA and
Livingston, LA) and the Virgo site (in Pisa, Italy), and a
four-detector configuration that adds the proposed LIGO-India detector
(in Chitradurga, KA).  For simplicity, we assume each detector to be
operating at the Advanced LIGO sensitivity.

For a multi-detector network, the noise-weighted inner products
\eqref{innerProduct} combine linearly so long as the noise is
uncorrelated between detectors, allowing us to use the above
formalism with an additional summation over interferometers in the network.  We integrate the inner product
from a lower-frequency cutoff of $20\text{Hz}$ to the
innermost-stable-circular orbit of the systems, which for
a non-spinning binary is a function only of the total mass:

\begin{equation}
  \pi f_{\text{ISCO}} = \frac{1}{6^{3/2}M_{tot}}.
  \label{ISCOFrequency}
\end{equation}

We define the signal-to-noise ratio (SNR) of a gravitational wave in a
single detector as
\begin{equation}
  \text{SNR} \equiv \frac{4}{\sigma} \int^{\infty}_{0}\frac{|
    \tilde{s}(f)\tilde{h}(f)|}{S_{n}(f)}df,
  \label{formalSNR}
\end{equation}
where $\tilde s(f)$ and $\tilde{h}^{*}(f)$ are
the frequency-domain data and template, respectively, and the
normalization $\sigma$ is given by
\begin{equation}
  \sigma^2 = 4\int^{\infty}_{0}\frac{| \tilde{h}(f)|^2}{S_n(f)}df.
  \label{SNRnorm}
\end{equation}

\noindent When dealing with a network of gravitational-wave detectors, the SNRs
of the individual detectors add in quadrature.  That is, the network
SNR for a detection is

\begin{equation}
\text{SNR}_{\text{network}} = \sqrt{\sum_i \text{SNR}_{i}^2}
\label{SNRnetwork}
\end{equation}

\noindent where $\text{SNR}_i$ is the SNR, given by \eqref{formalSNR}, of
the $i^{\text{th}}$ detector.

For this study, we consider four separate populations of BNS systems,
with component mass combinations of $1M_{\odot}/1M_{\odot}$,
$1.4M_{\odot}/1.4M_{\odot}$, $1M_{\odot}/2.5M_{\odot}$, and
$2.5M_{\odot}/2.5M_{\odot}$.  Each population consisted of 40 signals
distributed randomly in sky location, polarization, inclination,
time-of-arrival, and coalescence phase.  The luminosity distance, $D$,
was selected to yield a signal strength of $\text{SNR}_{network}=20$
for each source in each network configuration. 
 While unphysical from an astrophysics perspective, this
choice allows us to explore the parameter estimation in the context of
loud but plausible detection candidates.  Signals much stronger than $\text{SNR}=20$ will most likely be rare.  Signals much lower will be at the threshold for detection, where details of the instrument noise play a more commanding role in any inferences which will be drawn from the observations, making it impossible to meaningfully quantify ``typical'' parameter estimation accuracies until the detectors begin operating.

Finally, note that the PSD that defines the inner
product \eqref{innerProduct} is simply the time-averaged sensitivity
of a given detector to a specific frequency. Under ideal circumstances any stretch of data should contain a specific noise
realization drawn from a Gaussian colored by the PSD however, in practice, non-Gaussian noise events are occasionally present in the data.  While we await the completion of the advanced detectors, we are without a
realistic model including the actual instrument performance, and cannot accurately
simulate gravitational-wave data to the point of making complete predictions about
parameter estimation capabilities. Because of
this, we elect to focus only on the part of the simulation over which we have control by
setting $n(t) \equiv 0$. Thus, what is reported here are strictly the statistical errors due to the
flexibility of the waveforms and the expected sensitivity of the
advanced era detectors while suppressing any effects of random detector noise.  Consequently, what we recover with the MCMC should be interpreted as
the posterior distribution function for each source averaged over all
possible (Gaussian) noise realizations, assuming we have the correct
PSD and that the signal is sufficiently powerful for the linearized-signal
approximation to be valid.  By selecting a source population with an
 artificial distribution in SNR, we ensure that the ``zero-noise'' as average uncertainty approximation is valid.
 We discuss further the validity of this approach in Appendix \ref{noiseSection}. 

\section{Results}
\label{resultsSection} 

\begin{figure*}[t!]
\centering \includegraphics[trim=7cm 0cm 0cm 0cm,
  clip=true,scale=0.58]{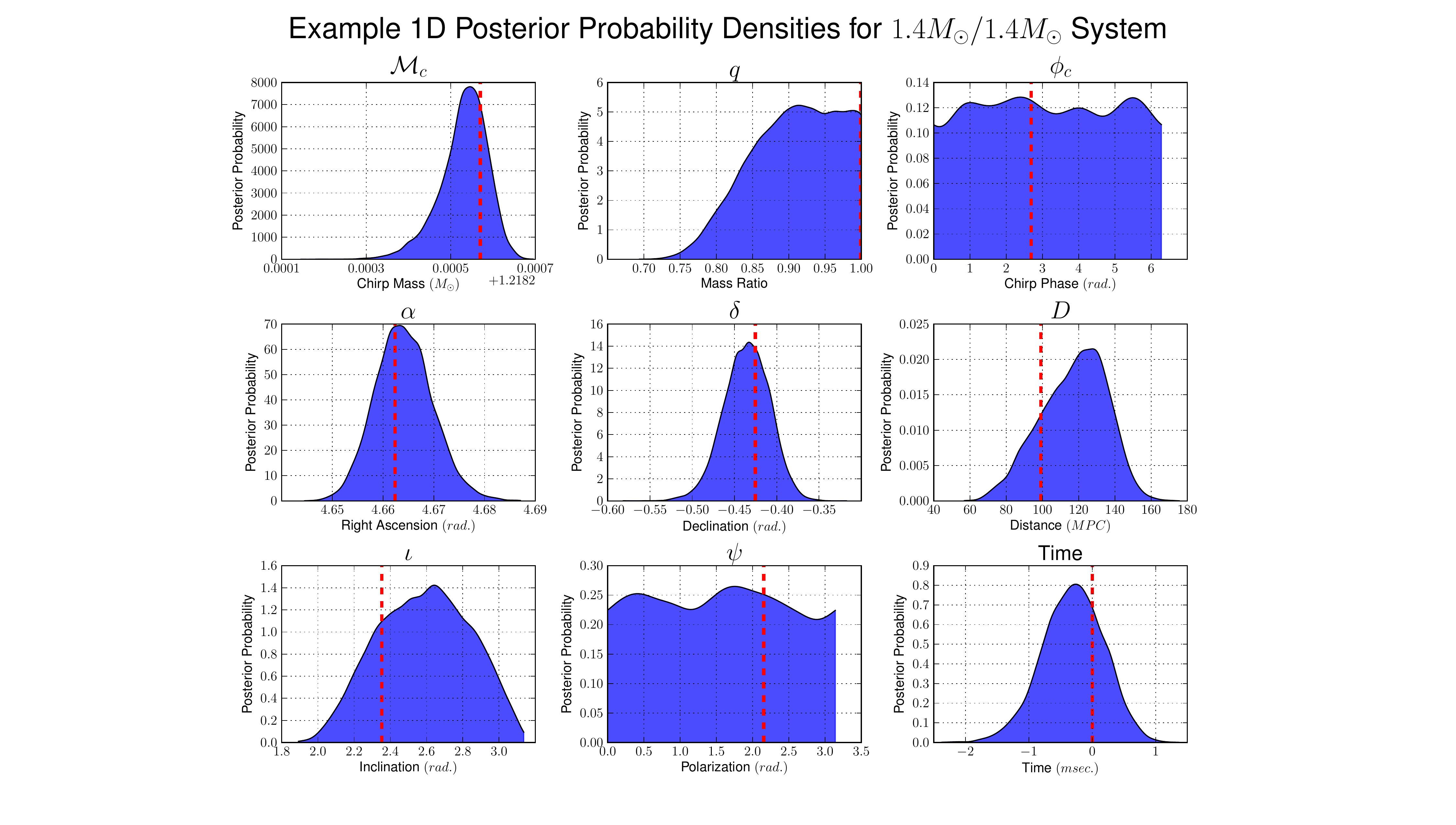}
\caption{\label{9dPDF} Marginalized 1D  posterior probability densities
  of a typical $1.4M_{\odot}/1.4M_{\odot}$ BNS
  system.  We have plotted each of the 9 parameters for our
  non-spinning BNS problem as stated in section \ref{parameterSection}.
    Note how, even for parameters with excellent recovery, the peak of
  the 1D Gaussian is displaced from the injected value (in dashed
  red).  This is not due to a systematic bias, but is caused by the
  marginalization of a single dimension from the full 9D posterior
  space.  To better see this effect, compare the 1D PDF for
  $\mathcal{M}_{c}$ and $q$ with the 2D marginalized PDF in
  Figure~\ref{1414masses}.  The plots here represent the Gaussian kernel density
  estimator of each PDF.  The plot for $q$ was 
  computed by reflecting the highest $10\%$ of samples across the $q=1$ boundary, in order
  for the smoothed plot to agree with the binned histograms.
  }
\end{figure*}

Of the nine parameters in the domain of the waveform, only six are
particularly physically interesting: the masses of the two binaries,
$M_1$ and $M_2$, the orbital inclination, $\iota$, the angular
position on the sky, $\alpha$ and $\delta$, and the luminosity
distance of the source, $D$.  While the coalescence phase $\phi_0$,
the coalescence time $t_c$, and the wave polarization $\psi$ must be
included in any parameter estimation of the waveform, they do not
encode any information of particular astrophysical
interest.

In Figure~\ref{9dPDF}, we provide an example of the nine,
1-dimensional marginalized posterior probability density functions
recovered from a single $1.4M_{\odot}/1.4M_{\odot}$ BNS system.  These
PDFs are representative of the type of results that will be produced by
parameter estimation studies in the advanced detector era.  Notice
that the peak of several parameters, including the chirp mass,
$\chmass$, appears to be displaced from the true values in dashed red.  This
effect is due to gradients in the priors on the mass parameters and distance, and the reduction of the 9-dimensional PDF to a series of
marginalized 1-dimensional PDFs.  For instance, the 1-dimensional PDF
for chirp mass is marginalized via
\begin{equation}
p(\chmass | s) = \int _{\thpara \setminus \chmass} p(\thpara |
s)~d(\thpara \setminus \chmass)
\label{mcMarginalization}
\end{equation}
where the notation $\thpara \setminus \chmass$ implies all parameter
of \eqref{parameterspace} except $\chmass$.  Other parameter and
higher-dimensional marginalizations follow a similar convention.  In
practice, the MCMC samples make this integral trivial: since the
samples are distributed according to the posterior,
\eqref{mcMarginalization} can be ``computed'' by simply histogramming the chain entries of a single parameter, implicitly calculating a
Monte-Carlo integral over all other parameters.

 We display marginalized 1D distributions after smoothing with a Kernel Density Estimator (KDE).
At a prior boundary (e.g. q=1) the KDE is artificially depleted because there is no support from beyond the edge in parameter space.  To rectify this it is customary to reflect some of the points across the boundary, so that the KDE resembles the histogram.  This estimator is unbiased as long as the first derivative of the prior is zero at the boundary. We elected to use 10\% of the points, as it was the minimum number required to get the smooth KDE to match the binned histograms.

\subsection{Mass Parameters}
\label{massSection}

\begin{figure}[h!]
  \centering \includegraphics[trim=1cm 0cm 2cm 0cm,
    clip=false,scale=0.63]{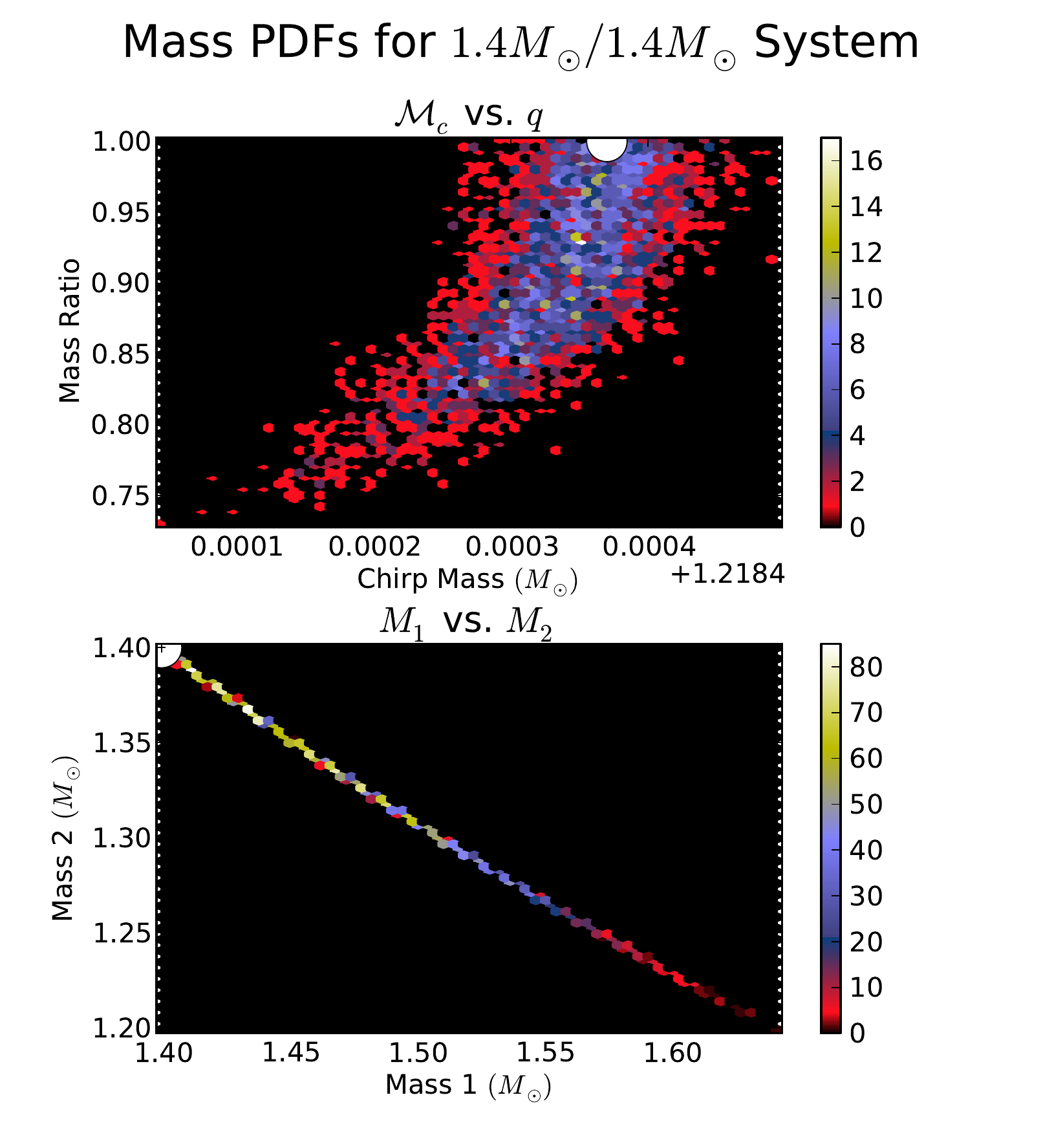}
 \caption{2D marginalized posterior probability density functions for
   the mass parameters recovered in typical a
   $1.4M_{\odot}/1.4M_{\odot}$ system.  The posteriors are plotted in
   terms of parameters used in the waveform, chirp mass
   ($\mathcal{M}_C$) and the mass ratio ($q$), and in the individual
   component masses of the binary ($M_1$ and $M_2$).  In the
   $\chmass$-$q$ space, the posterior would almost resemble a Gaussian if not
   for the limitation of $q \leq 1$.  The presence of the $q$ cutoff
   and the convention that $M_1 \geq M_2$ inform the non-Gaussian
   features present.  When projected as 1D marginalized posteriors,
   the component masses resemble the posterior PDFs shown in Figure
   \ref{metaMassPDFs}.}
  \label{1414masses}
\end{figure}

\begin{figure}[h!]
  \centering \includegraphics[trim=1cm 0cm 2cm 0cm,
    clip=false,scale=0.63]{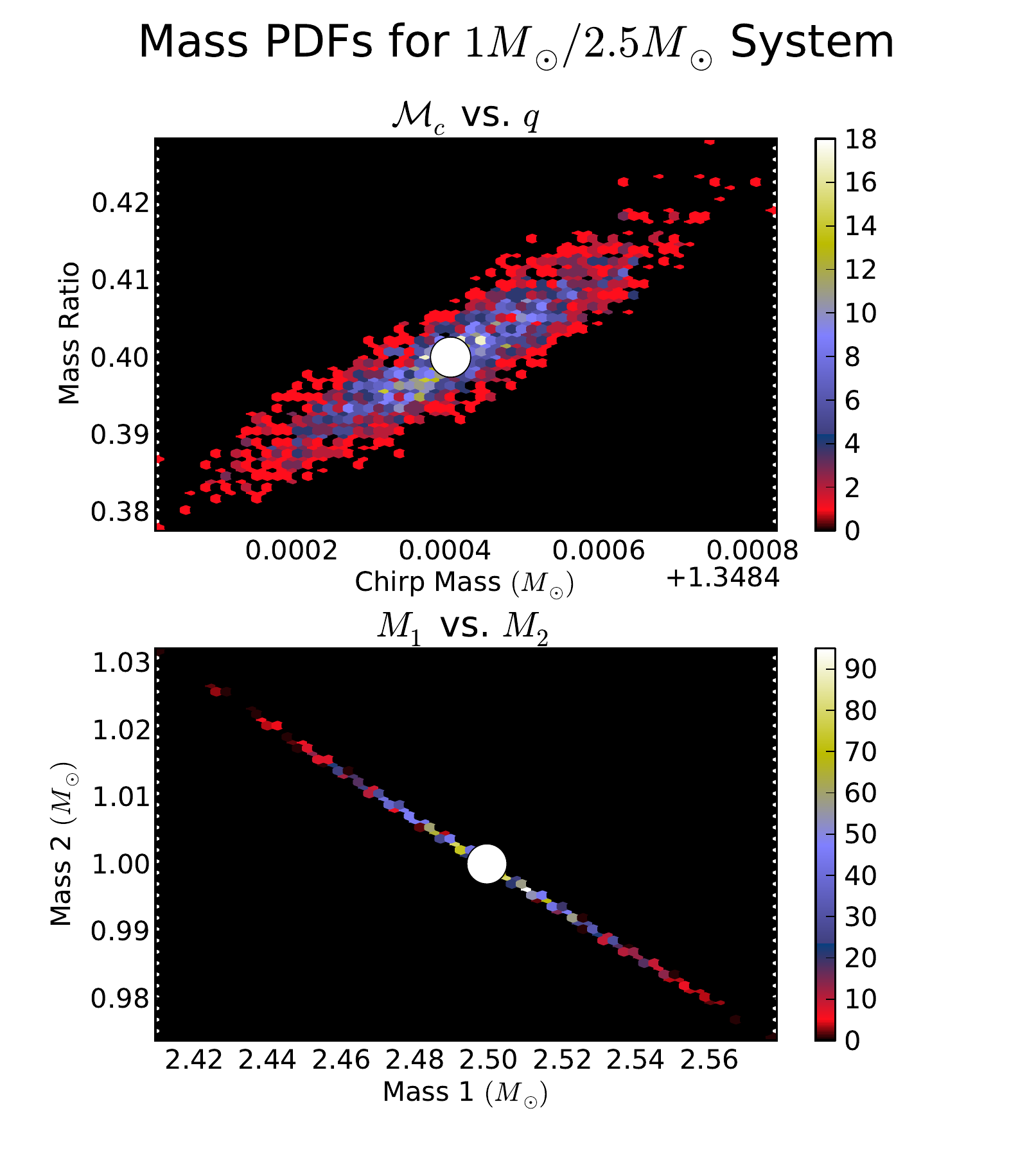}
 \caption{Similar to Figure~\ref{1414masses}, but for a typical
   $1M_{\odot}/2.5M_{\odot}$ system.  The unequal mass ratio displaces
   the posterior PDFs from the $q \leq 1$ boundary present in the equal-mass
   case, yielding a Gaussian PDF in both the $\chmass$-$q$ and
   $M_1$-$M_2$ spaces.}
  \label{125masses}
\end{figure}

\begin{figure*}[ht!]
  \centering \includegraphics[trim=3cm 0cm 2cm 0cm,
    clip=false,scale=0.75]{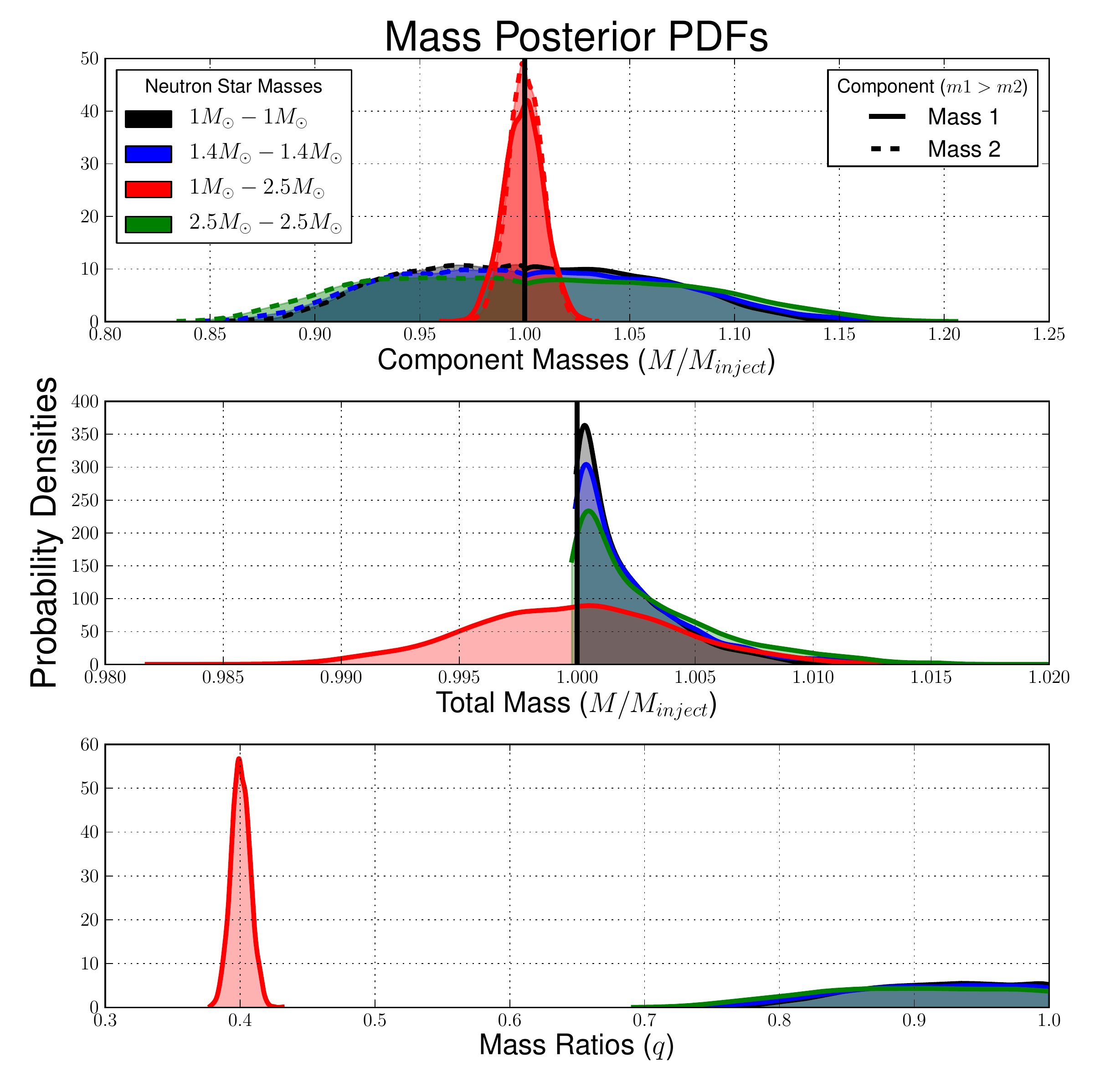}
 \caption{Mass PDFs for the four BNS systems, averaged over each of
   the 40 injections.  The average is reported since, in practice,
   there was quantitatively no difference between the recovered PDFs
   for different systems with identical masses and SNRs.  Systems with
   identical mass ratios but different total masses have
   quantitatively identical PDFs when normalized to the total
   mass.  Similarly to Fig. \ref{9dPDF}, we plot the Gaussian kernel density
   estimates, and reflect $10\%$ of the points across the $q = 1$ and $M_1 = M_2$ boundaries (for the three equal-mass cases).}
  \label{metaMassPDFs}
\end{figure*}

Of the nine variables in our parameter space \eqref{parameterspace},
the mass parameters, $\chmass$ and $q$, or, correspondingly, $M_1$ and
$M_2$ are the most astrophysically interesting.  The ability of
Advanced LIGO/Virgo to construct a population of BNS masses will be
one of the more useful and immediate applications of gravitational-wave astronomy.

If we sort our results by each system's mass parameters, we find virtually no difference between the mass PDFs of injections with the same intrinsic parameters.  More simply, at a given SNR the accuracy to which LIGO/Virgo can measure the mass parameters of a (non-spinning) BNS system is independent of the sky-location and orientation of the binary.
This can be understood if we recall that the mass
parameters are the only two which directly affect the phase evolution of the
waveform \eqref{phase}.  Therefore, as long as the source's mass parameters are equivalent, and the injected SNR is the same for each source, the amount of recoverable information pertaining to the phase of the waveform is identical.
The only noticeable difference will come from
the specific realization of noise produced by the detector, which we
address in Appendix \ref{noiseSection}.



In Figures~\ref{1414masses} and~\ref{125masses}, we show the
marginalized 2D posterior PDFs of our mass parameters for prototypical
equal-mass and unequal-mass binaries.  We include the PDF in both the
$\chmass$-$q$ space (relevant for the waveform and the MCMC
algorithm), and the more physically interesting component mass space
($M_1$-$M_2$).  Although only the $1.4M_{\odot}/1.4M_{\odot}$ system
is included in Figure \ref{1414masses}, the PDF is qualitatively
identical to the other equal mass cases, modulo a scaling factor.

To condense the results into a single figure of merit, we average the
1-dimensional mass PDFs into a single posterior probability for the
system in Figure~\ref{metaMassPDFs}.  Notice how, when averaged and
normalized to the injected values, the recovery of the component
masses depends only on the mass ratio.  Furthermore, for systems with
equal component masses, the posterior barely extends beyond 15\% of
the injected values.  If one assumes that the threshold between black
holes and neutron stars lies at approximately $3M_{\odot}$, then it
might be naively assumed that the mass recovery would allow one to
discriminate between black holes and neutron stars.

In practice, there is a physical complication to this claim: we have
neglected the spin of the binaries, which will be highly correlated to
the masses.  This coupling means that, in the case where the component masses have non-negligible spins
  parallel to the orbital angular momentum of the binary, a higher-massed spinning system can produce waveforms very similar to a non-spinning,
low-mass binary.  This effect makes it extremely difficult to discern between, for example, 
 non-spinning neutron stars and spin-aligned, stellar mass black holes
\citep{Baird2013,Hannam2013}.  However, the situation is not hopeless:
if the spins are misaligned, the spin vectors will couple to the
orientation of the binary (encoded in the three angles $\phi_0$,
$\iota$, and $\psi$) via relativistic precession.  It remains to be
seen if using fully spinning waveforms will make it possible for
Advanced\ LIGO/Virgo to discern binary neutron stars from their
spinning black hole counterparts.

In Table \ref{ciTableIntrinsic}, we list the 68\% and 95\% credible
intervals for the mass parameters.  We find that the component masses
for equal mass systems can be isolated to between 6.4\% and 9\% (10.3\% and 15\%)
fractional uncertainty at the 68\% (95\%) credible interval.  This value
drops to less than 1.9\% (3.7\%) for the components of the unequal mass systems.
Again, this neglects the effects of spin which can substantially
increase these values.

\subsection{Inclination and Distance}
\label{idSection}

\begin{figure}[h!]
  \centering \includegraphics[trim=0cm 0cm 0cm 0cm,
    clip=true,scale=0.75]{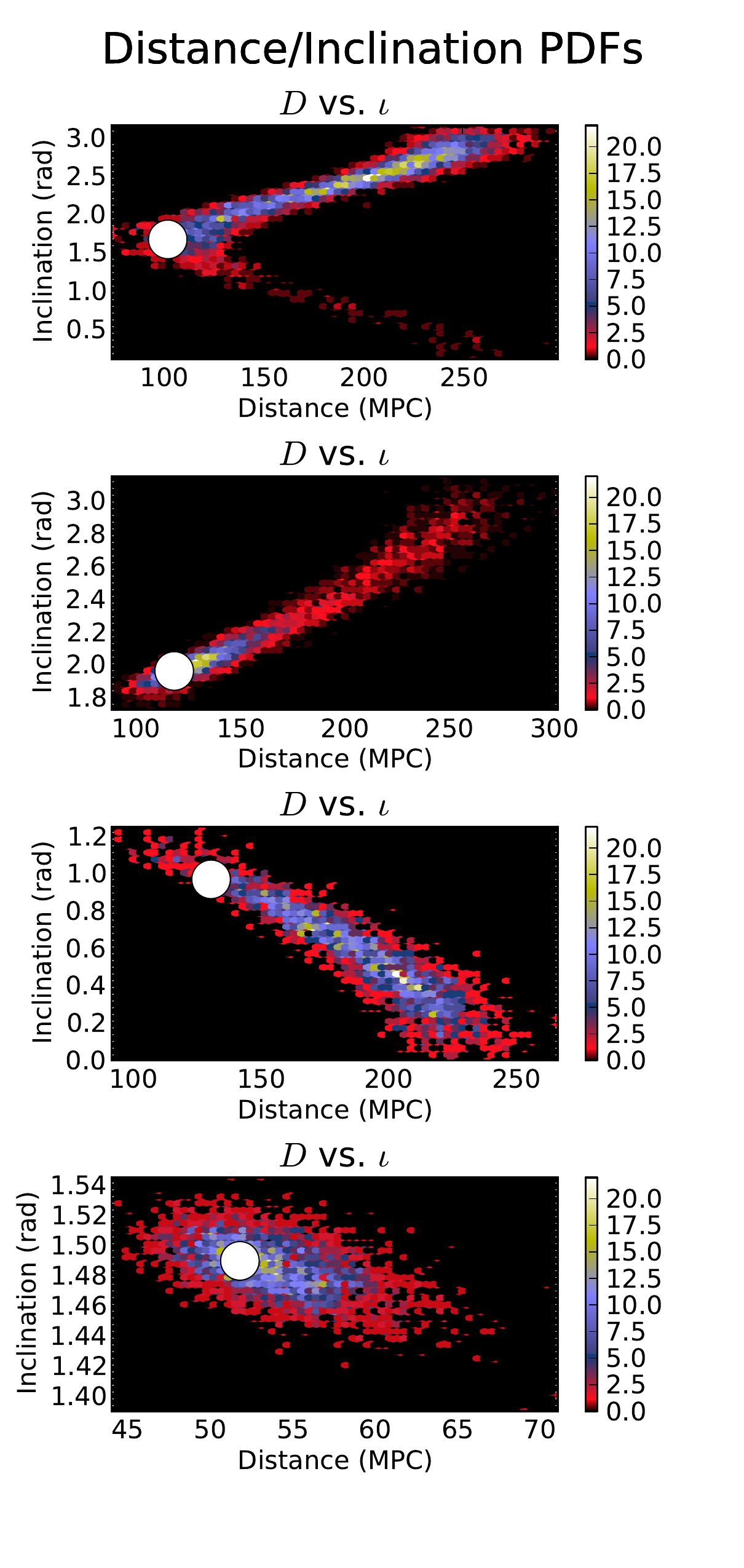}
 \caption{Typical marginalized 2D mass PDFs for luminosity distance
   $D$ and orbital inclination $\iota$. Notice the bimodal
   distribution on the first PDF, that can occur when a system is
   detected nearly ``edge-on'' ($\iota \approx \pi/2$).  In general,
   the degeneracy between distance and inclination forces the PDF into
   one of these two paths when the system is not edge-on, as seen in
   the second and third PDFs.}
 \label{distIotaPDF}
\end{figure}

Binary neutron star systems (along with neutron star/black hole
systems) are one of the best candidates for progenitors of short-hard
gamma-ray bursts \citep[and references therein]{Nakar2007}.  After
merger, the remnant is believed to emit the burst along the axis of
orbital angular momentum, making the inclination of the binary system
of particular interest to gamma-ray astronomy
\citep{LSCGRB2010,Corsi2012}.  The inclination is detected as a
relative amplitude difference between the two gravitational-wave
polarizations, such that to lowest order:

\begin{align}
h_+(f) &= \frac{1+\cos^2(\iota)}{2 D} \tilde{h}(f) \nonumber
\\ h_\times(f) &= i \frac{\cos(\iota)}{D}\tilde{h}(f) .
\end{align}

\noindent It should be apparent that the luminosity distance $D$ and
the inclination $\iota$ can be highly correlated in any parameter
estimation recovery.

Given this degeneracy, it should come as no surprise that the 2D
marginalized posteriors of distance and inclination are the broadest
of the six physically interesting parameters.  Four typical 2D PDFs
are presented in Figure~\ref{distIotaPDF}.  Notice the bimodal
uncertainty present in the top 2D PDF along the $\iota$ axis, due to
the similarity between the evaluated likelihoods at $\iota$ and $\iota
+ \pi/2$.  As the majority of the information extracted via parameter
estimation is from the phase of the signal, the recovery of the
posterior in $D$-$\iota$ space will be limited, even during the
Advanced LIGO/Virgo era; however, if a gravitational-wave signal is
matched to a detected SGRB counterpart (which may occur at the rate of
$\sim 3 yr^{-1}$ in Advanced LIGO/Virgo, according to
\cite{Metzger2013}), the optical information about the orbital
inclination will provide an additional constraint on the D-$\iota$
space, vastly improving the luminosity distance recovery quoted here.  These results
qualitatively agree with those of \cite{Nissanke2011}, who studied in detail
MCMC estimation of $D$-$\iota$ measurements from gravitational-wave detectors with 
coincident GRB detections.

In Table \ref{ciTableExtrinsic}, we list the averaged 68\% and 95\% credible
intervals for distance and inclination.  Notice that the uncertainties
on distance can be extreme, from tens to hundreds of Mpc.  When
discussing the inclination angle, we elected to use $|\cos(\iota)|$,
as this maps the occasional bimodal structure into a single PDF, and
observing a compact merger at $\iota$ and $\iota + \pi/2$ should yield
identical physics.

\subsection{Sky Localization}
\label{skySection}

Unlike the mass parameters, the recovery of the position on the sky is
highly dependent on the location of the source with respect to the
detector network in question.  Much of the information about the sky
position of the source comes from the relative timing of the signals
in each detector.  For the HLV configuration there are two regions
in the sky which produce consistent delays in the arrival time of the GW signal between pairs of detectors.  That is, when using only time-of-arrival information,
there is equal support in the probability density functions around the true
location and the region of the sky found by reflecting the correct line of sight through the plane formed by the
three detectors.  See \cite{Fairhurst2011} for a global analysis of
time-of-arrival accuracy for various network configurations, including
those considered here.  See also \cite{Prospects2013} for a similar analysis,
applied to each stage of the advanced detector network.

In practice, however, additional information from the wave
polarization and the relative difference in SNR between individual
detectors can break this plane-reflection degeneracy, leading to a
bi-modal distribution with significantly more support in one mode of
the PDF.  By fitting for the sky location and the polarization
simultaneously, the MCMC can often identify the correct mode on the
sky.  For the four-detector configuration, this concern is irrelevant:
even with time-of-arrival data alone, the HLVI network can constrain
any source to a single mode on the sky.  Even then, there are still
locations on the sky in which two or more detectors are not sensitive
to the gravitational-wave signal, yielding distorted, non-Gaussian
PDFs.

\begin{figure*}[ht!]
  \centering \includegraphics[angle=0,scale=0.4, trim=5cm 2cm 3cm
    0cm]{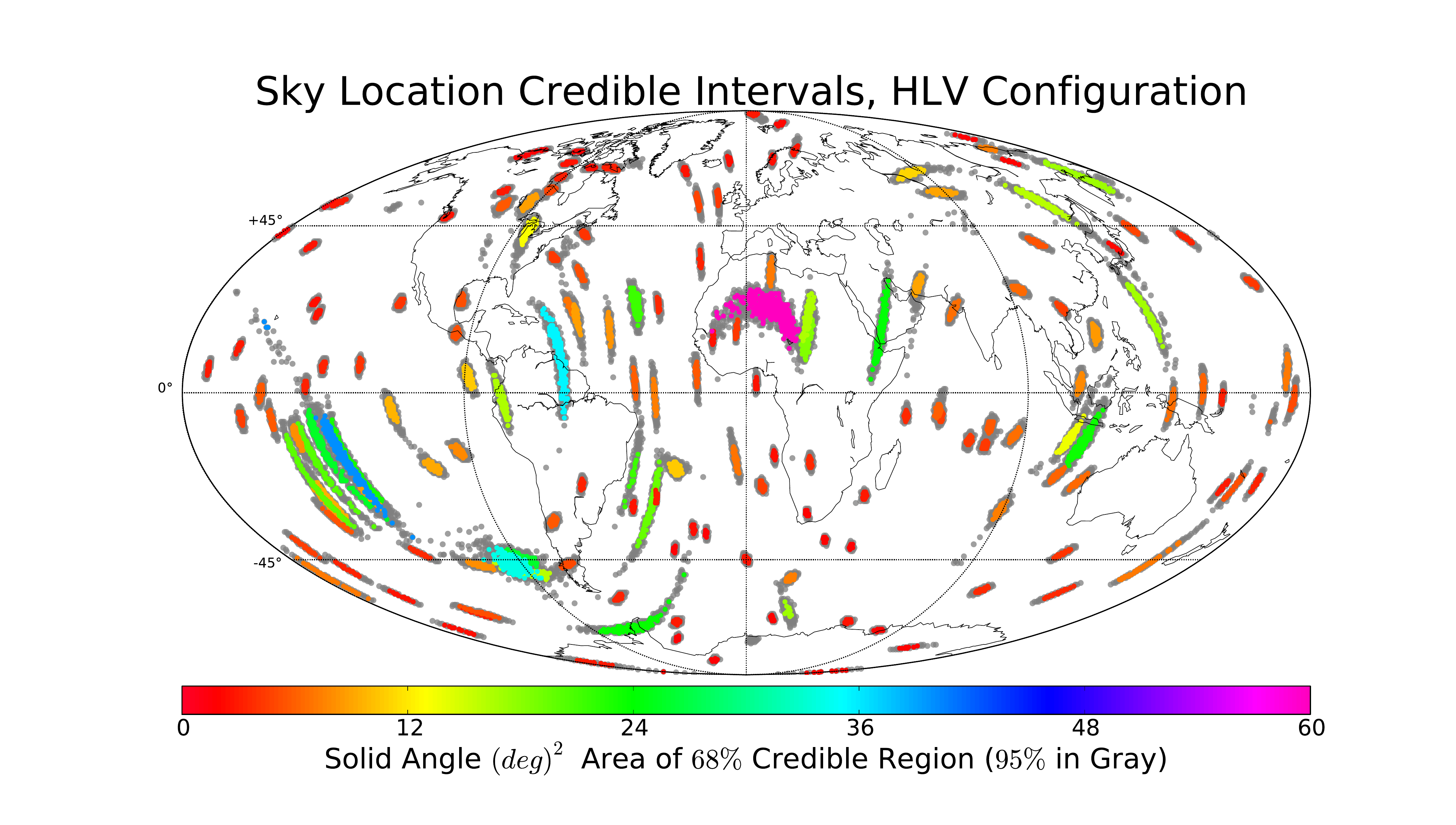}
 \caption{The uncertainties on the sky of 160 BNS systems in the HLV
   detector configuration.  Each region represents a single injection,
   with the colored central region representing the 68\% uncertainty
   region on the sphere, and the gray shade representing the 95\%
   uncertainty region.  The color scheme indicates the total solid
   angle size of the 68\% region.  Note the similar shape of the
   uncertainty regions at particular points; this is due to the
   specific pattern of sensitivity over the sky for the three-detector network.}
 \label{2525SkyLocHLV}
\end{figure*}

\begin{figure*}[ht!]
  \centering \includegraphics[angle=0,scale=0.4, trim=5cm 2cm 3cm
    0cm]{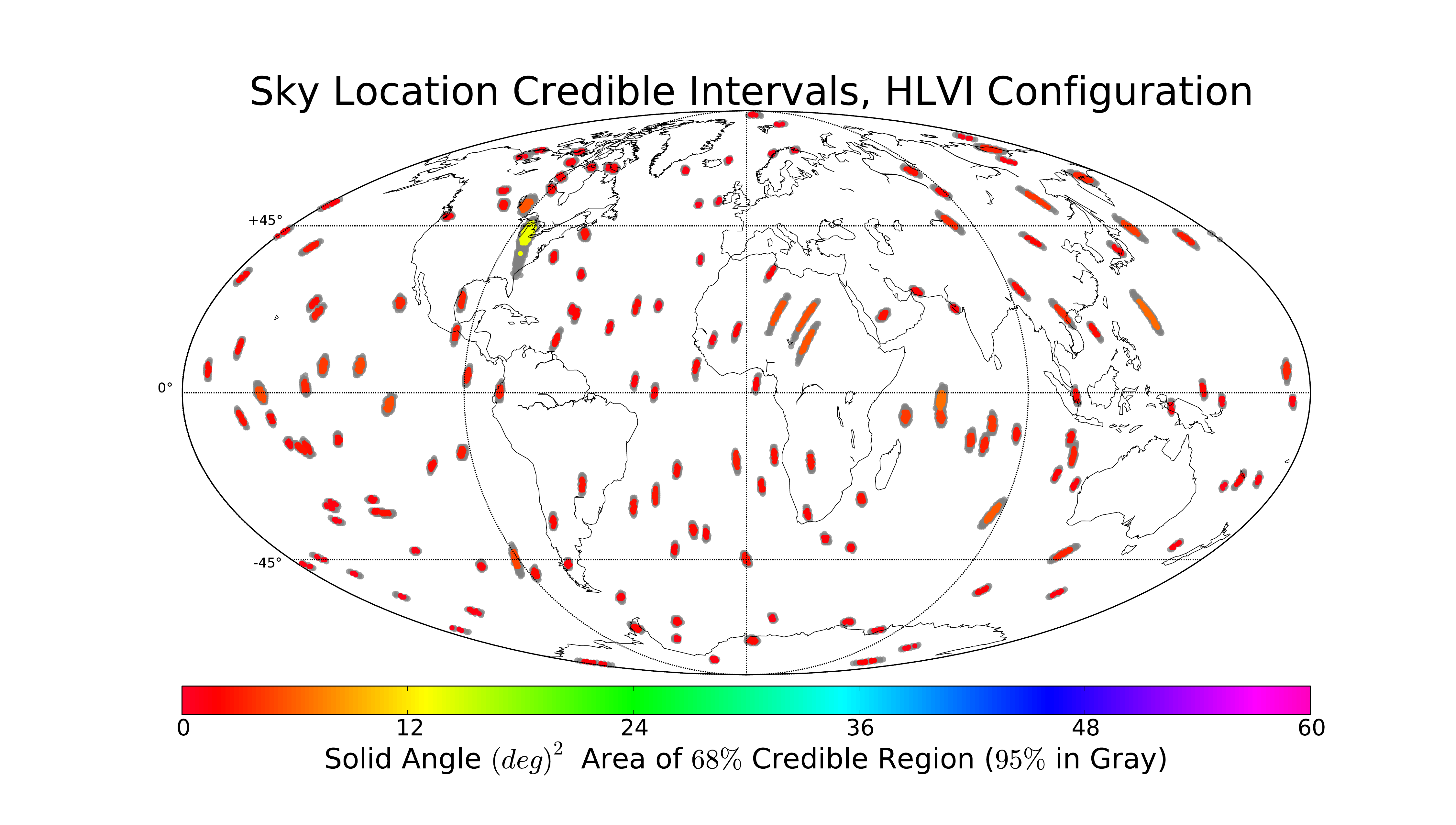}
 \caption{The same as Fig \ref{2525SkyLocHLV}, except for the HLVI
   detector configuration.  Note the substantially lower average
   uncertainties on the skies for the majority of the injections.
   Also note the lack of large, ``banana-shaped'' uncertainties that
   were recovered by the HLV configuration.  The two improvements are
   strictly due to the breaking of the plane degeneracy that
   is facilitated by the transition to a four-detector network, and not the additional SNR achieved by adding another detector to the network.  The SNR was held at 20 for both figures.}
 \label{2525SkyLocHLVI}
\end{figure*}

The sky location uncertainties for HLV and HLVI are shown in Figures
\ref{2525SkyLocHLV} and \ref{2525SkyLocHLVI} respectively.  We show
all four mass bins together, since in practice the mass of the signals
has little effect on the recovery of the sky location for the
considered BNS systems.  Only the location on the sky, the network
configuration, and the other extrinsic parameters were found to be
relevant.  In particular, notice the increase in efficiency between
the HLV configuration and the HLVI configuration.  
The network SNR is held fixed for each configuration so the improvements seen here are strictly due to extra information provided by the additional interferometer, not the increased SNR a fourth detector would provide for sources at a fixed distance.

In the HLV
configuration, there exist points on the sky in which the signal was
injected near the plane of the detector network.  This causes the
elongated, ``banana-shaped'' PDFs seen in Figure \ref{2525SkyLocHLV}.
In the HLVI configuration, the plane is no longer relevant, and there
are substantially fewer regions in which only two detectors see a
sufficiently strong signal.  For the HLV configuration, we find that
 all of the signals are recovered with
a solid angle uncertainty of less than 64 $\mathrm{deg}^2$ (136 $\mathrm{deg}^2$)
at the 68\% (95\%) credible interval, while the HLVI configuration 
recovers all signals to less than 14 $\mathrm{deg}^2$ (45 $\mathrm{deg}^2$).
The median quantitative results for the two network configurations show a similar
 benefit with the addition of LIGO India, and are again reported in Table \ref{ciTableExtrinsic}. 

\begin{figure}[ht!]
  \centering \includegraphics[angle=0,scale=0.58, trim=0.5cm 1cm 0cm 0cm]{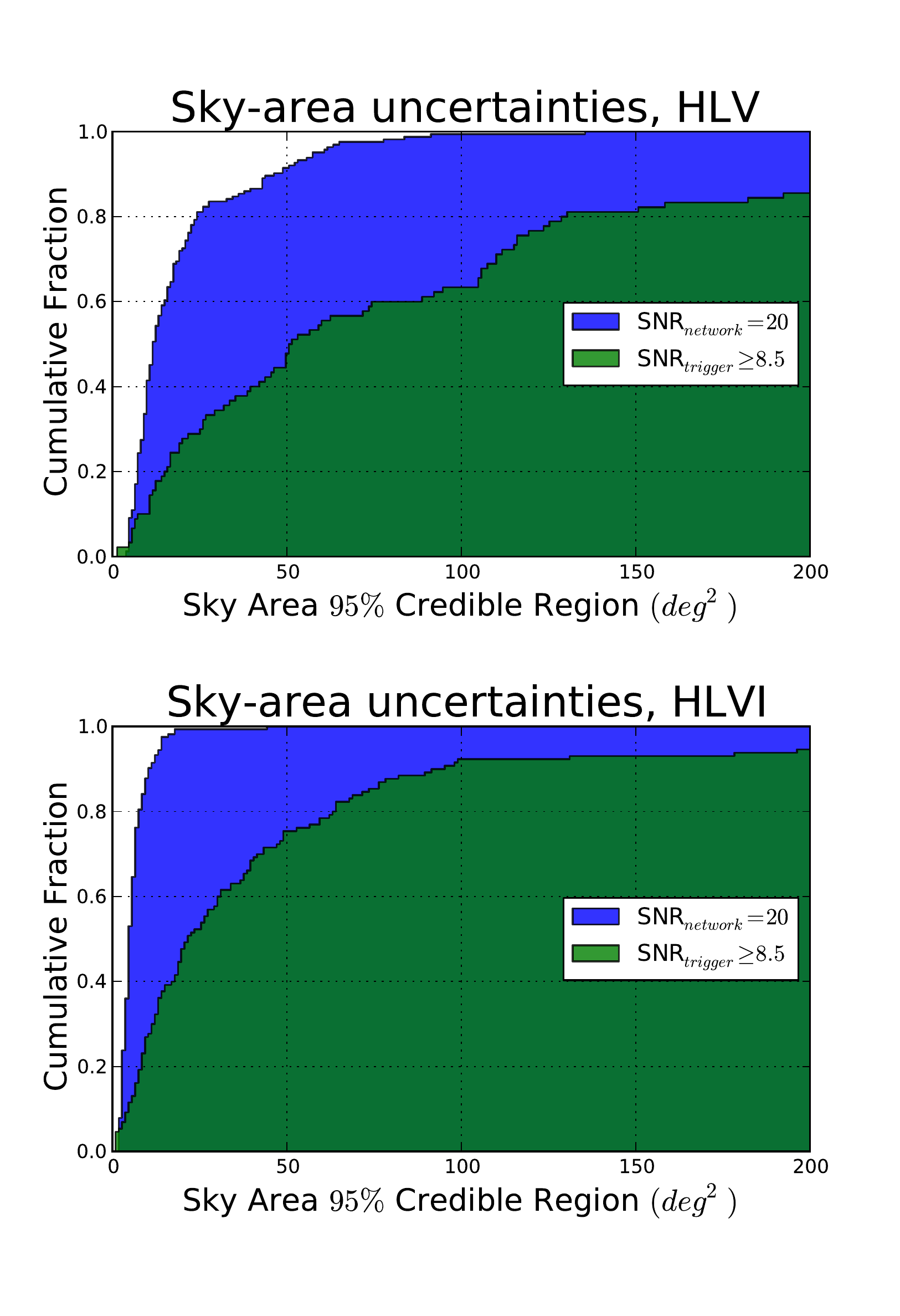}

 \caption{The cumulative fraction of sources found within $deg^2$ at the 95\% credible
 interval.  This plot contains results from two separate studies: the current study, in
 which only strong detection candidates are considered ($\text{SNR}_{network}=20$), 
and \cite{Nissanke2013}, which employs a separate MCMC code on a
 realistic source population (where $\text{SNR}_{trigger} \geq 8.5$ is employed to determine
 parameter estimation sources).  The strong detection candidates are constrained
 to substantially smaller solid angles on the sky, as is expected for such ``gold-plated''
 events, while both studies show a substantial decrease in sky-area uncertainties for the HLVI configuration.}
 \label{compSky}
\end{figure}

Previous studies have considered the increase in sensitivity from the addition of the 
LIGO India site.  \cite{Nissanke2013} studied the decrease in
sky-location uncertainties obtained by the inclusion of the LIGO India site
into the advanced detector network.  Their study also employed an MCMC sampling technique 
 (a modified version of CosmoMC),
to determine the average parameter estimation uncertainties for a realistic distribution
of sources.  As the current study is interested in the average uncertainties
of strong sources at $\text{SNR}_{network}=20$, ignoring the effects of less viable 
detection candidates, it is useful to compare the $95\%$ credible intervals
between the two studies.  The cumulative distributions of sky-area uncertainties
for both network configurations in each study are plotted in Fig. \ref{compSky}.

The difference in uncertainties between the two studies is illustrative.  \cite{Nissanke2013}
considered a realistic distribution of sources motivated by a local galaxy catalogue for
 sources where $D\leq 200\text{Mpc}$, and homogeneously distributed in co-moving
 volume for sources where $D > 200\text{Mpc}$.  The ``detection candidates'' which underwent parameter estimation were limited to the signals
 where $\text{SNR}_{network} \geq 8.5$.   The SNR threshold for a detection is a simplification to the actual detection statistic used for BNS searches.
   The current study considers a source
 population isotropically distributed on the sky, but at distances such that each signal
 represents a relatively high precision case for parameter estimation.   Fig.\ \ref{compSky} compares
 the sky-area uncertainties that will be available to those considering ``good 
 gravitational-wave events'' (e.g. precision source localization), and the 
 uncertainties that one can gather by considering all gravitational-wave candidates with 
 $\text{SNR}_{network}\geq 8.5$ (e.g. electromagnetic follow-up).  As expected, both source
 populations show a dramatic decrease in parameter estimation uncertainties with the
 addition of the LIGO India site to the advanced detector network.  This holds true even when
 considering the increased number of nominal sources available to the HLVI network 
 configuration in the Nissanke population, where the expansion of the network detection 
 horizon due to LIGO India increases the number of sources to 130, versus 90 in 
 the HLV configuration.  This trend is similar to that observed in \cite{Veitch2012}, contrasting
 the originally proposed four-detector configuration (with two co-located Hanford detectors) to 
 the HLVI configuration.

\section{Credible Intervals}
\label{ciSection}

When quoting parameter estimation results, it is often convenient to
reduce the full parameter space to credible intervals about single
marginalized parameters..  To that end, we state the averaged 68\% and 95\% credible
intervals about the single parameters for each of the three
parameter pairs of interest.  These were already plotted for the sky
locations in Figures~\ref{2525SkyLocHLV} and \ref{2525SkyLocHLVI}.

There are several different ways of computing the width of a credible
interval in this setup.  If one considers the points in the MCMC as
random draws from the true posterior, then the $\alpha$-level credible
interval can be computed by simply ordering the points, removing
$N(1-\alpha )/2$ points from both sides of the posterior samples
symmetrically, and measuring the width of the remaining points.  While
this procedure can prove hazardous for multi-modal distributions, it
is straightforward and reliable for single-peaked distributions as
reported here.

In Table~\ref{ciTableIntrinsic}, we list the mean 68\% and 95\% credible
intervals recovered for the four mass configurations in both network
configurations. This essentially quantifies the widths of the
posteriors plotted throughout Section~\ref{resultsSection}.  The
purpose of Tables~\ref{ciTableIntrinsic} and \ref{ciTableExtrinsic} is
to provide a quantitative and quotable source for studies seeking to
determine how well a physical question about BNS systems can be
answered with the Advanced LIGO/Virgo network.

Given the previous reliance of parameter estimation studies on the Fisher
Information Matrix, it is informative to compare the credible intervals quoted here 
to previously reported values in the literature.  We recompute the credible intervals 
of the chirp mass and the symmetric mass ratio ($\eta \equiv m_1 m_2 / M_{tot}^2$)
at the 68\% confidence level ($1\sigma$) for the $1.4M_{\odot}/1.4M_{\odot}$ system presented 
here.  We then compare the findings of the MCMC to the Fisher matrix predictions using
an identical network configuration (HLV), noise curve, and waveform model.  
We compute the Fisher matrix uncertainties 
using a code previously described in \cite{Inadequacies}.  We find that the MCMC credible
 intervals can vary significantly from
the estimates of the Fisher matrix, with the Fisher matrix underestimating the 
fractional uncertainties by as much as a factor of two.  See Table~\ref{FIMvsMCMC}.  This 
result agrees with the conventional wisdom surrounding the Fisher Matrix accuracy,
as well as results where the Fisher Matrix is computed to higher order \citep{Vitale2010}
   This disagreement, while 
   unsurprising, emphasizes the importance of employing the full parameter estimation
    machinery of an MCMC when making physically relevant claims.

\begin{table*}[h!]
\centering
\caption{Median 68\% and 95\% credible intervals for intrinsic parameters for
  each of the four systems considered.  We report the credible
  intervals of quantities measured, as well as the component masses
  and total mass.  Although the results for the HLV and HLVI
  configurations are quantitatively identical, we report them
  separately for consistency.}
  \tabcolsep=0.11cm
      {\renewcommand{\arraystretch}{1.3} 
\begin{tabular}{lcccccccccccccccccccc}

\\ HLV\\

 \hline\hline System & \vline & \multicolumn{3}{c}{$\Delta \chmass / \chmass$} &
\vline & \multicolumn{3}{c}{$\Delta M_1 / M_1$} & \vline & \multicolumn{3}{c}{$\Delta M_2 / M_2$} & \vline &
\multicolumn{3}{c}{$\Delta M_{tot}/M_{tot}$} & \vline & \multicolumn{3}{c}{$\Delta q$}\\ \hline
Credible Level & \vline & 68\% & \vline & 95\% & \vline & 68\% & \vline & 95\% & \vline & 68\% & \vline & 95\% & \vline & 68\% & \vline & 95\% & \vline & 68\% & \vline & 95\% \\
\hline\hline

$1M_{\odot}-1M_{\odot}$ & \vline &$0.00497\%$ & \vline &$0.0104\%$ & \vline & $7.17\%$ & \vline &$11.9\%$ & \vline & $6.39\%$ & \vline &$10.3\%$ & \vline & $0.643\%$ & \vline &$1.25\%$ & \vline & $0.123$ & \vline &$0.197$\\\hline$1.4M_{\odot}-1.4M_{\odot}$ & \vline &$0.00883\%$ & \vline &$0.0188\%$ & \vline & $7.77\%$ & \vline &$13\%$ & \vline & $6.87\%$ & \vline &$11.1\%$ & \vline & $0.746\%$ & \vline &$1.47\%$ & \vline & $0.132$ & \vline &$0.212$\\\hline$1M_{\odot}-2.5M_{\odot}$ & \vline &$0.0176\%$ & \vline &$0.0355\%$ & \vline & $1.86\%$ & \vline &$3.74\%$ & \vline & $1.59\%$ & \vline &$3.23\%$ & \vline & $1.48\%$ & \vline &$2.99\%$ & \vline & $0.0138$ & \vline &$0.028$\\\hline$2.5M_{\odot}-2.5M_{\odot}$ & \vline &$0.0246\%$ & \vline &$0.0522\%$ & \vline & $9.02\%$ & \vline &$15\%$ & \vline & $7.82\%$ & \vline &$12.6\%$ & \vline & $1.01\%$ & \vline &$1.94\%$ & \vline & $0.149$ & \vline &$0.239$\\

\hline
\hline

\\HLVI\\

 \hline\hline System & \vline & \multicolumn{3}{c}{$\Delta \chmass / \chmass$} &
\vline & \multicolumn{3}{c}{$\Delta M_1 / M_1$} & \vline & \multicolumn{3}{c}{$\Delta M_2 / M_2$} & \vline &
\multicolumn{3}{c}{$\Delta M_{tot}/M_{tot}$} & \vline & \multicolumn{3}{c}{$\Delta q$}\\ \hline
Credible Level & \vline & 68\% & \vline & 95\% & \vline & 68\% & \vline & 95\% & \vline & 68\% & \vline & 95\% & \vline & 68\% & \vline & 95\% & \vline & 68\% & \vline & 95\% \\
\hline\hline

$1M_{\odot}-1M_{\odot}$ & \vline &$0.00497\%$ & \vline &$0.0106\%$ & \vline & $7.15\%$ & \vline &$11.9\%$ & \vline & $6.38\%$ & \vline &$10.4\%$ & \vline & $0.646\%$ & \vline &$1.27\%$ & \vline & $0.123$ & \vline &$0.198$\\\hline$1.4M_{\odot}-1.4M_{\odot}$ & \vline &$0.00884\%$ & \vline &$0.0188\%$ & \vline & $7.67\%$ & \vline &$12.8\%$ & \vline & $6.79\%$ & \vline &$11.1\%$ & \vline & $0.733\%$ & \vline &$1.46\%$ & \vline & $0.13$ & \vline &$0.211$\\\hline$1M_{\odot}-2.5M_{\odot}$ & \vline &$0.0176\%$ & \vline &$0.0352\%$ & \vline & $1.85\%$ & \vline &$3.72\%$ & \vline & $1.59\%$ & \vline &$3.2\%$ & \vline & $1.47\%$ & \vline &$2.96\%$ & \vline & $0.0137$ & \vline &$0.0277$\\\hline$2.5M_{\odot}-2.5M_{\odot}$ & \vline &$0.0243\%$ & \vline &$0.0515\%$ & \vline & $9.03\%$ & \vline &$14.9\%$ & \vline & $7.84\%$ & \vline &$12.6\%$ & \vline & $0.998\%$ & \vline &$1.92\%$ & \vline & $0.149$ & \vline &$0.238$\\

\hline
\hline

\end{tabular}}
\label{ciTableIntrinsic}
\end{table*}

\begin{table*}[h!]
\centering
\caption{Median 68\% and 95\% credible intervals of extrinsic parameters for
  each of the four systems considered.  As expected, there exists a
  substantial improvement in the sky localization capabilities of the
  four-detector HLVI configuration over the three-detector HLV
  configuration.  Note that the solid-angle sky-location credible
  intervals, $\Delta\Omega$, are calculated directly on the 2D sphere,
  not by combining the $\alpha$ and $\delta$ uncertainties.}

  \tabcolsep=0.11cm
      {\renewcommand{\arraystretch}{1.3} 
\begin{tabular}{lcccccccccccccccccccc}

\\ HLV\\

\hline\hline System & \vline & \multicolumn{3}{c}{$\Delta D$ $(mpc)$} & \vline &
\multicolumn{3}{c}{$\Delta |\cos(\iota)|$} & \vline & \multicolumn{3}{c}{$\Delta \alpha$ $(deg)$} & \vline &
\multicolumn{3}{c}{$\Delta \delta$ $(deg)$} & \vline & \multicolumn{3}{c}{$\Delta\Omega$
$(deg^2)$}\\\hline
Credible Level & \vline & 68\% & \vline & 95\% & \vline & 68\% & \vline & 95\% & \vline & 68\% & \vline & 95\% & \vline & 68\% & \vline & 95\% & \vline & 68\% & \vline & 95\% \\
 \hline\hline 

$1M_{\odot}-1M_{\odot}$ & \vline &$49.4$ & \vline &$89.9$ & \vline & $0.323$ & \vline &$0.611$ & \vline & $1.73$ & \vline &$4.09$ & \vline & $2.51$ & \vline &$5.52$ & \vline & $5.12$ & \vline &$13.5$\\\hline$1.4M_{\odot}-1.4M_{\odot}$ & \vline &$61.4$ & \vline &$107$ & \vline & $0.314$ & \vline &$0.588$ & \vline & $2.63$ & \vline &$5.42$ & \vline & $2.53$ & \vline &$5.27$ & \vline & $4.12$ & \vline &$11.2$\\\hline$1M_{\odot}-2.5M_{\odot}$ & \vline &$68.8$ & \vline &$127$ & \vline & $0.31$ & \vline &$0.549$ & \vline & $2.41$ & \vline &$4.6$ & \vline & $2.77$ & \vline &$6.2$ & \vline & $4.37$ & \vline &$12.1$\\\hline$2.5M_{\odot}-2.5M_{\odot}$ & \vline &$116$ & \vline &$198$ & \vline & $0.349$ & \vline &$0.613$ & \vline & $1.75$ & \vline &$4.51$ & \vline & $2.42$ & \vline &$5.01$ & \vline & $4.62$ & \vline &$12$\\

\hline\hline

\\HLVI\\ 
\hline\hline System & \vline & \multicolumn{3}{c}{$\Delta D$ $(mpc)$} & \vline &
\multicolumn{3}{c}{$\Delta |\cos(\iota)|$} & \vline & \multicolumn{3}{c}{$\Delta \alpha$ $(deg)$} & \vline &
\multicolumn{3}{c}{$\Delta \delta$ $(deg)$} & \vline & \multicolumn{3}{c}{$\Delta\Omega$
$(deg^2)$}\\\hline
Credible Level & \vline & 68\% & \vline & 95\% & \vline & 68\% & \vline & 95\% & \vline & 68\% & \vline & 95\% & \vline & 68\% & \vline & 95\% & \vline & 68\% & \vline & 95\% \\
 \hline\hline 
$1M_{\odot}-1M_{\odot}$ & \vline &$42.7$ & \vline &$76.2$ & \vline & $0.267$ & \vline &$0.455$ & \vline & $1.13$ & \vline &$2.25$ & \vline & $1.48$ & \vline &$3.01$ & \vline & $1.87$ & \vline &$5.37$\\\hline$1.4M_{\odot}-1.4M_{\odot}$ & \vline &$66.6$ & \vline &$121$ & \vline & $0.285$ & \vline &$0.509$ & \vline & $1.27$ & \vline &$2.48$ & \vline & $1.49$ & \vline &$3.01$ & \vline & $2$ & \vline &$5.12$\\\hline$1M_{\odot}-2.5M_{\odot}$ & \vline &$73.7$ & \vline &$130$ & \vline & $0.297$ & \vline &$0.499$ & \vline & $1.29$ & \vline &$2.42$ & \vline & $1.42$ & \vline &$2.89$ & \vline & $1.75$ & \vline &$4.87$\\\hline$2.5M_{\odot}-2.5M_{\odot}$ & \vline &$120$ & \vline &$213$ & \vline & $0.301$ & \vline &$0.517$ & \vline & $1.18$ & \vline &$2.34$ & \vline & $1.55$ & \vline &$3.14$ & \vline & $2.25$ & \vline &$5.99$\\ \hline

\end{tabular}}
\label{ciTableExtrinsic}
\end{table*}

\begin{table}[h!]
\centering
\caption{68\% Credible intervals versus the $1\sigma$ Fisher matrix confidence intervals for mass parameters in each system in the HLV configuration.}
  \tabcolsep=0.11cm
    {\renewcommand{\arraystretch}{1.3} 
\begin{tabular}{lcccccccc}

\hline\hline Parameter & \vline & \multicolumn{3}{c}{$\Delta\mathcal{M}_c / \mathcal{M}_c$} & \vline & \multicolumn{3}{c}{$\Delta \eta / \eta$} \\ \hline \hline

Estimate & \vline & MCMC & \vline & FIM & \vline & MCMC & \vline & FIM \\ \hline

$1M_{\odot}-1M_{\odot}$ & \vline &  0.00497\% & \vline & 0.00396\% & \vline & 0.645\% & \vline & 0.704\% \\
$1.4M_{\odot}-1.4M_{\odot}$ & \vline &  0.00890\% & \vline & 0.00743\% & \vline & 0.745\% & \vline & 0.866\% \\
$1M_{\odot}-2.5M_{\odot}$& \vline &  0.0176\% & \vline & 0.00880\% & \vline & 1.475\% & \vline & 0.862\% \\
$2.5M_{\odot}-2.5M_{\odot}$& \vline &  0.0245\% & \vline & 0.0221\% & \vline & 1.000\% & \vline & 1.299\% \\

\hline\hline

\end{tabular}}
\label{FIMvsMCMC}
\end{table}

\section{Conclusion}
\label{conclusionSection}

In this paper, we performed an MCMC parameter estimation analysis on
the recoverability of basic information about binary neutron stars,
using two projected versions of the Advanced LIGO/Virgo network.  We
focused on the recovery of the two masses, the luminosity distance,
orbital inclination, and the sky location, as these are the six basic
parameters of physical interest to the problem.  The signals were injected at a high, 
but not unrealistic, ${\rm SNR}=20$ in order to create a reference similar to those found in previous
Fisher Matrix studies while remaining in a regime where the ``$n(t) = 0$ as average'' simplification is justified.  
The simulated signals comprehensively covered sky-location, orientation, component mass, and mass ratios for plausible binary neutron star systems, while neglecting spin, corrections due to finite size effects, orbital eccentricity, and other higher-order modifications to the gravitational wave signal.

 Despite the high SNR of our simulated signals, Fisher matrix results are not adequate to characterize the parameter estimation capabilities of advanced detectors.  The Fisher matrix approximation assumes the likelihood distribution is a multivariate Gaussian.  In contrast, using MCMC to sample the posterior distribution function allows us to relax any assumptions about the functional form of the posterior (apart from the implicit dependence on templates and the PSD common to any similar study).  Given the highly non-Gaussian nature of the recovered posteriors, even in the absence of a simulated noise realization, (see Figures \ref{9dPDF}, \ref{1414masses}, \ref{metaMassPDFs}-\ref{2525SkyLocHLVI} ), we conclude that it is critical for a complete 
understanding of the gravitational-wave BNS parameter estimation problem to use Bayesian sampling techniques.  

The quantitative results reported here represent the average statistical error for ${\rm SNR}=20$ detections -- ``loud'' but not unrealistic for  Advanced LIGO/Virgo
observations -- assuming well-modeled waveforms and detectors which achieve their design sensitivity.
For the mass parameters we found that, neglecting the effects of spin, the component masses can be
constrained to within 9\% (15\%) of their true value to a credible level of
68\% (95\%) for systems with equal-mass neutron stars.  This value drops below 1.9\% (3.7\%) for systems with an asymmetric masses.
It was also found that the fractional uncertainties for equal-mass binary systems are similar at equal SNR.
Only when the masses are unequal are the fractional errors effected.
These results were summarized in Table \ref{ciTableIntrinsic}. { Our study is the first 
to comprehensively quantify neutron star mass measurements in Advanced LIGO/VIRGO  via an MCMC for a range of component masses and mass ratios.} 

We also reported on the ability of the two network configurations to
constrain the luminosity distance and orbital inclination.  For
distance, it was found that the uncertainties will average anywhere from
43 to 120 MPC at 68\% credible levels, and from 76 to 213 MPC at 96\% credible levels,
 making the uncertainties 
larger than the luminosity distances themselves in many cases.  Furthermore, it was
found that the cosine of the orbital inclination can be constrained to
within 0.35 (0.61) at the 68\% (95\%) level on average,
 suggesting that Advanced LIGO/Virgo will not
be able to offer constraining information on GRB beaming angles in coincidence
with electromagnetic observations.  Coincidence detections will still be possible, but
the orbital orientation information from gravitational waves will 
not provide astrophysically relevant constrains.

Finally, we reported the ability of advanced networks to constrain the
sky location of strong BNS signals.  It was found that the three-detector
configuration, consisting of the Washington and Louisiana LIGO sites
plus the Italian Virgo site, was able to constrain all signals within
64 $\mathrm{deg}^2$ (136 $\mathrm{deg}^2$) on the sky at the 68\% (95\%) 
credible level, with
an average median 68\% (95\%) credible interval of 
$4.6$ $\mathrm{deg}^2$ (12.2 $\mathrm{deg}^2$).
Meanwhile, the four-detector configuration, consisting of the
three-detector sites plus a LIGO India detector, was able to localize
all the sky locations to within 14 $\mathrm{deg}^2$ (45 $\mathrm{deg}^2$) on the sky at the 68\% (95\%)
credible interval, with an average median 68\% (95\%) credible interval of 
2 $\mathrm{deg}^2$ $(5.3$ $\mathrm{deg}^2)$.  These high-SNR results
were then compared to similar results for a more realistic, source distribution from
\cite{Nissanke2013}.

It should be noted that there are two distinct types of systematic
error, highly relevant to the gravitational-wave parameter estimation
problem, that we have not addressed in this study.  First, we have
studied the parameter estimation uncertainties under the assumption
that the waveform template we use to recover the signal template
exactly matches the fully relativistic waveforms nature provides.  In
practice, these waveforms are only approximations to the fully
general-relativistic physics required to solve the problem.  See
\cite{BuonannoWaveform} for a better description of the systematic uncertainties
present in the most common waveform families.  Additionally, there are
several astrophysical assumptions that can potentially contribute to
systematic uncertainties in the waveforms, such as the neutron-star
equation of state, possible modifications to General Relativity,
eccentricity, etc.

Secondly, we have performed our study under idealized
detector-noise conditions.  In practice the noise levels of Advanced
LIGO and Advanced Virgo can vary with time and contain occasional
excursions which are highly non-Gaussian.  Unfortunately, there is no
reliable way to predict the sort of non-Gaussian detector glitches and
instrumentation effects that will arise in any advanced
gravitational-wave detector.  The realization of noise will be a
major factor in the deflection of signal PDFs from the idealized cases
presented here.

In this study, we have also neglected the effects of spin in the
parameter space, electing to focus on the absolute basic parameters
that will be measured routinely in the Advanced Detector era.  Given
the high degree of coupling between the spin and mass of objects
in the gravitational-wave parameter space, it remains unclear if the
mass measurement alone will be sufficient to distinguish non-spinning
neutron stars from highly-spinning low-mass black holes.  Future work
will explore this potential mass/spin degeneracy, including the
effects of orbital precession, with the aim of definitively answering
this question.

\pagebreak

\section{Acknowledgements}

The authors would like to thank Samaya Nissanke, John Veitch, and Neil Cornish for useful discussions.  We would also like to thank Samaya Nissanke for the 95\% credible regions
quoted in Fig. \ref{compSky}. CR and BF were supported by NSF GRFP Fellowships, award DGE-0824162.  All authors were also partially or fully supported by NSF Gravitational Physics grant PHY-0969820, PI: VK. Lastly, VK is grateful to the hospitality of the Aspen Center for Physics while she worked on this study. 

\bibliographystyle{apj} \bibliography{paper}{}

\appendix

\section{Markov Chain Monte-Carlo}
\label{appendixMCMC}

  The MCMC sampler included in \texttt{LALInference} employs a Metropolis-Hastings sampling algorithm , which is described as follows:

\begin{enumerate}
\item Pick an initial point in the parameter space
  ($\boldsymbol{\theta_{\text{old}}}$), and then propose a random
  ``jump'' to a new set of waveform parameters,
  $\boldsymbol{\theta_{\text{new}}}$.  The jump follows the
  proposal distribution $q\left(
  \boldsymbol{\theta_{\text{new}}} | \boldsymbol{\theta_{\text{old}}}
  \right)$.
\item Calculate the posterior probability,
  $p(\boldsymbol{\theta_{\text{new}}}|s)$, of the new parameters using
  \eqref{likelihood} and \eqref{posterior}.
\item Accept the new parameters with probability
  \begin{equation}
    p_\mathrm{accept} = \min \left[ 1,
      \frac{p(\boldsymbol{\theta_{\text{new}}}|s)
        q\left(\boldsymbol{\theta_{\text{old}}} |
        \boldsymbol{\theta_{\text{new}}}
        \right)}{p(\boldsymbol{\theta_{\text{old}}}|s)
        q\left(\boldsymbol{\theta_{\text{new}}} |
        \boldsymbol{\theta_{\text{old}}} \right)} \right].
  \end{equation}
  If the new parameters are accepted, record
  $\boldsymbol{\theta_\text{new}}$ and repeat with
  $\boldsymbol{\theta_\text{new}} \rightarrow
  \boldsymbol{\theta_\text{old}}$; otherwise, record
  $\boldsymbol{\theta_\text{old}}$, and repeat.
\end{enumerate} 

The above procedure is designed to record a chain of samples whose
distribution is $p\left(\thpara|s\right)$.  By drawing a sufficient
($\sim1000$) number of effectively independent samples from the
posterior, the chains traces out the functional form of the posterior,
gathering more samples from regions with high posterior probability
density.  Depending on the proposal distribution, $q$, the convergence
(mixing) of the chain may be rapid or slow.  We employ multiple
optimization techniques, including both specially-crafted jump
proposals ($q$) and parallel tempering, to ensure adequate mixing of
the Markov Chains throughout our parameter space.  Both samplers were
tuned and developed during the last science run of the Initial
LIGO/Virgo network.  A description of the parameter estimation
capabilities of these two samplers with respect to real interferometer
data, as well as a more detailed description of the algorithms and
checks for convergence, can be found in \cite{S6PE}.

\section{Gravitational-Waveform Model}
\label{appendixWaveform}

We use a frequency-domain waveform accurate up to $3.5$ post-Newtonian
(pN) order in phase and 3 pN order in amplitude of the lowest ($l=m=2$) spatial mode.
  We restrict
ourselves to quasi-circular, non-spinning waveforms as a simplifying
assumption.  The standard form of our waveform model, known as the
\textit{TaylorF2} approximant, is calculated via the stationary-phase
approximation.  In this setup, the gravitational-wave amplitude is
given by
\begin{equation}
\tilde{h}(f) = a(t_f) e^{i \psi(f)},
\label{amplitude}
\end{equation}
where $a(t_f)$ is the amplitude evaluated at a stationary-phase
reference point, which to lowest order takes the form $a(t_f) \propto
f^{-7/6} \chmass^{5/6}\Theta(\text{angle})/D$, where $D$ is the
luminosity distance of the binary, and $\psi(f)$ is the pN phase.
$\Theta(\text{angle})$ is a function of the orbital orientation with
respect to the detector network in terms of the sky position, orbital
inclination, and the wave polarization.  In addition to the total
mass, $M_{tot}\equiv M_1+M_2$, it is convenient to work with the mass
ratio and chirp mass, defined in \ref{parameterspace}.
  Note that most gravitational-wave literature instead uses
  the \textit{symmetric mass ratio}, defined as $\eta \equiv M_1 M_2 /
  M^2$.  We elect to use $q$ as it is more physically intuitive, and
  because it avoids an integrable singularity that can appear for
  equal-mass systems when employing a prior on $p(\eta)$.  
     By convention, $M_1 \geq M_2$.  The stationary phase then
becomes an expansion in the Newtonian orbital velocity, $v=(\pi
M_{tot} f)^{1/3}$,
\begin{equation}
\psi(f) = 2 \pi f t_c - \phi_0 + \frac{\pi}{4} +
\frac{3}{128}\left(\frac{M_{tot}}{\chmass}\right)^{5/3}\sum^{n}_{k=0}\alpha_{k}v^{k-5}
\label{phase}
\end{equation}

\noindent where the $\alpha_{k}$ coefficients are taken from the pN
expansion to order $n/2$.  See \cite{BuonannoWaveform} for a
description and comparison of different waveform families.  The terms
$t_c$ and $\phi_0$ in equation (\ref{phase}) are constants of
integration, referred to as the chirp time and coalescence phase,
respectively.

\section{Zero-Noise versus Varied Noise Realizations}
\label{noiseSection}

\begin{figure}[b!]
  \centering \includegraphics[trim=2cm 0cm 0cm 0cm,
    clip=true,scale=0.52]{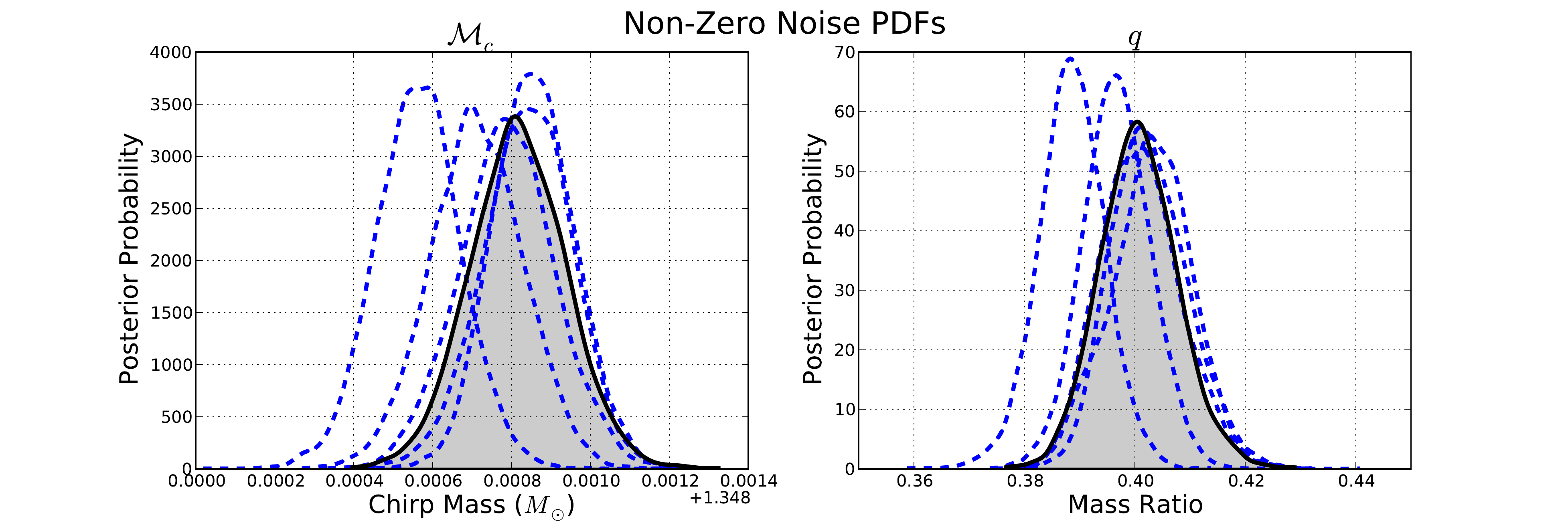}
 \caption{The effects of a non-zero noise realization on the recovered
   PDF for five different Gaussian realizations of the Advanced LIGO
   noise curve.  Each blue-dashed PDF represents the recovery of the same
   $1M_{\odot}/2.5M_{\odot}$ signal in a different noise curve, picked
   at random from the Gaussian colored noise defined by the Advanced
   LIGO power spectral density, while the gray-shaded curve is the
   zero-noise PDF.  Notice how each curve is a Gaussian PDF with different mean but similar 
   width from the zero-noise PDF. This is to be expected as the zero-noise mean is identical to the 
   average to first order in $1/\text{SNR}$, while the zero-noise standard deviation is identical to the average 
   standard-deviation
    up to $1/\text{SNR}^3$. 
   This only holds true for glitch-free data, which is an unrealistic
   idealization when compared to real data.}
  \label{noisePDFs}
\end{figure}

     We have performed the current analysis on zero-noise injections for
two reasons.  First, the results of a zero-noise analysis are
similar to those that would be achieved by averaging the results of
multiple identical injections in different Gaussian noise
realizations.  For a unique Gaussian realization of the Advanced LIGO
noise curve, the realization causes the maximum likelihood of the
posterior probability distribution to be translated away from the true
value. These displacements tend cancel in the frequentist average over noise
 realizations, ensuring that the mean uncertainties should be nearly identical to 
   the uncertainty drawn from the zero-noise runs.  This can be seen in 
   \cite{Vallisneri}, equations (73) and (74).
  By expanding the posterior mean and the posterior variance as series in $1/\text{SNR}$,
   it can be seen that the 
  $n(t) = 0$ posterior mean is identical to the frequentist average 
  over noise realizations of the posterior mean to first order in $1/\text{SNR}$.  In similar
  fashion, the $n(t) = 0$ posterior variance is identical to the noise-realization frequentist 
  average up to third order in $1/\text{SNR}$.  As the current study operates in a high-SNR 
  regime, the zero-noise averages (particularly the averages of the posterior variances) are
   assumed to be valid.

     Previous studies have both employed and studied in detail the consequences of this approach.
 In particular, \cite{Nissanke2010} demonstrated that the $n(t)=0$ posterior is equal to the geometric mean over 
     specific noise realizations of the posterior, while \cite{Sampson2013}
performed a numerical experiment to 
     demonstrate the validity of the assumption.  We perform a similar numerical experiment here in which we demonstrate that the uncertainties from a $n(t)=0$ posterior are equivalent to the frequentist
     average over uncertainties from multiple noise realizations.   We injected a single
$1M_{\odot}/2.5M_{\odot}$ system, detected in HLVI, into 5 separate
realizations of Gaussian noise colored by the Advanced LIGO PSD, and
compared the results to the same MCMC recovery in the zero-noise case.
This example can be seen for one-dimensional marginalizations of
$\chmass$ and $q$ in Fig. \ref{noisePDFs}.  In effect, the ``real
answer'' that will be recovered is a single PDF with similar width to
the zero-noise PDF, but with the peak likelihood displaced from the
true value.

  Of course, this also relies on the assumption that the detector noise is Gaussian.  
  This is the second reason for employing the zero-noise approximation:  it is not feasible
to predict the characteristics of realistic
advanced detector noise.   While we can simulate a simple non-Gaussian excursion or errors in PSD estimation, there is an infinite set of possibilities and picking any one is too speculative prior to the advanced detectors are completed and collecting data.
However, as the current study is focused on the average uncertainties,
 and not the systematics of a single event (for which we must await a true detection), we feel the n=0 method most adequately satisfies requests from the community for quotable Òrule-of-thumbÓ estimates.
  
 Techniques which build
off of the theoretical progress made
in~\cite{Allen:2002jw,Rover:2011qd,Littenberg:2010gf} for including
glitches in the model for the data, and therefore relaxing the
assumptions about stationary and Gaussian noise, are currently under
investigation.

\end{document}